\documentclass[12pt,preprint]{aastex}
\usepackage{epsfig}
\usepackage{emulateapj5}

\shorttitle{The orientation of galaxies}

\shortauthors{Trujillo, Carretero \& Patiri}

\begin{document}

\title{Detection of the effect of cosmological large--scale structure on the
orientation of galaxies}

\author{Ignacio Trujillo\altaffilmark{1,2}, Conrado Carretero\altaffilmark{3}
and Santiago G. Patiri\altaffilmark{3}} 

\altaffiltext{1}{Max--Planck--Institut f\"ur Astronomie, K\"onigstuhl 17, D--69117 Heidelberg,
Germany}
\altaffiltext{2}{School of Physics \& Astronomy, University of Nottingham,
University Park, Nottingham, NG7 2RD, UK; ignacio.trujillo@nottingham.ac.uk}
\altaffiltext{3}{Instituto de Astrof\'{\i}sica de Canarias, Calle V\'{\i}a
L\'actea, E-38200, La Laguna, Tenerife, Spain; cch@iac.es, spatiri@iac.es}

\begin{abstract}

Galaxies are not distributed randomly throughout space but are instead arranged
in an intricate "cosmic web" of filaments and walls surrounding bubble-like
voids. There is still no compelling observational evidence of a link between the
structure  of the cosmic web and how galaxies form within it. However, such a
connection is expected on the basis of our understanding of the origin of galaxy
angular momentum: disk galaxies should be highly  inclined relative to the plane
defined by the large-scale structure  surrounding them. Using the two largest
galaxy redshift surveys currently in existence (2dFGRS and SDSS) we show at the
99.7\% confident level that these alignments do indeed exist: spiral galaxies
located on the shells of the largest cosmic voids have rotation axes that lie
preferentially on the void surface. 

\end{abstract}
\keywords{galaxies: spiral --- galaxies: structure --- 
galaxies: haloes  --- large--scale structure of the universe -- methods:
statistical --- dark matter}

\section{Introduction}

The spin of spiral galaxies is believed to be generated by tidal torques
operating in the early Universe on the primordial material destined to form a
galaxy (Peebles 1969; White 1984). A generic prediction of  this theory is the
existence of local correlations between galaxy rotation  axes and the
surrounding matter field (Barnes \& Efstathiou 1987; Porciani et al. 2002a,b;
Navarro, Abadi \& Steinmetz 2004, hereafter NAS04 ). Several observational
studies (Helou \& Salpeter 1982, Flin \& Godlowski 1986, Garrido et al. 1993,
NAS04) have looked for a preferential spatial orientation (or alignment) of
galaxy rotation axes with respect to their surrounding large-scale structures. 
These attempts have been focused on galaxies in the vicinity of the Milky-Way.
Our Galaxy and its neighbours are embedded in a slab, the supergalactic plane
(de Vaucouleurs 1953), that extends out to several tens of megaparsecs from our
Galaxy. Prior work has concentrated on detecting an excess of nearby galaxies
whose rotation axes were approximately perpendicular to the normal to the
supergalactic plane. The existence of such an alignment has been difficult to
establish, with claims of detection but also claims of no convincing evidence
for alignments (Dekel 1985). A number of factors complicate the above analysis.
For example, the direction of the rotation axis of a spiral galaxy requires
knowledge not only of the position angle and inclination but also which side of
the galaxy is closer to the observer. This latter information is known only for
the closest galaxies to our own. This uncertainty, however, can be avoided if
only edge-on or face-on galaxies are selected (NAS04). In this case, the spatial
orientation of the rotation axis lies in the sky plane (edge-on case) or points
exactly towards or away from us (face-on case). Nevertheless, this approach has
the disadvantage that it reduces significantly the sample size. A natural
solution would be to explore the orientation of edge-on and face-on spiral
galaxies in two-dimensional large-scale distributions of galaxies beyond the
supergalactic plane. To explore the alignment of the galaxies with respect to
their surrounding matter it is necessary to characterize the spatial orientation
of the cosmic sheet (or plane) at the position of the galaxy. This can be done
by estimating the direction of the minor axis of the sheet at that point. This
minor axis is, however, very difficult to measure in practice because of
redshift-space distortion: peculiar velocities not associated with the Hubble
flow that prevent us from deriving to obtain the true three dimensional
arrangement of the galaxies. Consequently, the detection and characterization of
large-scale sheet-like distributions of galaxies in the redshift-space is ill
defined.  In this paper we present an alternative way of exploring preferential
spatial orientations without resorting to the detection of planes.  We
concentrate our analysis on galaxies located on the shells of the largest cosmic
voids (r$_{void}$$>$10 h$^{-1}$ Mpc). We adopt a cosmological model with
H$_0$=100 h km s$^{-1}$ Mpc$^{-1}$, $\Omega_m$=0.3 and $\Omega_\Lambda$=0.7.

\section{Void Based Method and Sample Selection}

The gradient of the density field at points around large voids shows a strong
tendency to be aligned with the direction of the vector going from the centre of
the void to those points. So, any departure from isotropy of the distribution of
the angle between the galaxy plane and the local "sheet" shall be inherited by
the distribution of the angle $\theta$ between the galaxy "spin" $\vec{s}$ and
the vector joining the galaxy with the centre of the corresponding void
$\vec{r}$ (Fig. 1).  Thus,  $\theta$ can be easily determined as:

\begin{equation}
\theta=\arccos\left(\frac{\vec{s}\cdot\vec{r}}{|\vec{s}||\vec{r}|}\right)
\end{equation}

When the galaxy lies in the shell of a large void, the vector $\vec{r}$ can be
approximated by the vector joining the centre of the void to the centre of the
galaxy. If the position of the void centre is $\vec{r}_{void}$ and the position
vector of the galaxy is  $\vec{r}_{galaxy}$, then $\vec{r}$ is simply:
$\vec{r}$=$\vec{r}_{galaxy}$-$\vec{r}_{void}$. To evaluate the spin vector
$\vec{s}$  of a galaxy  we need to know its position angle, $\phi$ , its
semi-minor to semi-major axis ratio, b/a, and its coordinates
($\alpha$,$\delta$). We define the inclination angle as: $\zeta=\arcsin(b/a)$.

In our definition, $\phi$ increases anticlockwise from North to East in the
plane of the sky. So, if we consider a galaxy characterized by angles ($\alpha$,
$\delta$, $\phi$, $\zeta$)=(0,0,0,0)  (i.e. an edge-on galaxy along the z axis),
the spin of this galaxy should be $\vec{s}_0$=(0,1,0). Therefore, the spin
$\vec{s}$ of a generic galaxy can be computed as the following product:
$\vec{s}$=M($\alpha$, $\delta$, $\phi$, $\zeta$)$\vec{s}_0$ where M($\alpha$,
$\delta$, $\phi$, $\zeta$) is the transformation matrix which considers all the
rotations for the four angles. If we denote R$_q$($\phi$) as the rotation matrix
of angle $\phi$ around the axis q, then we have that:  M($\alpha$, $\delta$,
$\phi$, $\zeta$)=R$_z$(-$\alpha$)R$_y$(-$\delta$)R$_x$($\phi$)R$_z$($\zeta$).
Introducing the explicit expressions for the rotation matrices and substituting
the value of M, we find:

\begin{eqnarray}
s_x=\cos\alpha\cos\delta\sin\zeta+\cos\zeta\left(\sin\phi\cos\alpha\sin\delta-\cos\phi\sin\alpha
\right)\\
s_y=\sin\alpha\cos\delta\sin\zeta+\cos\zeta\left(\sin\phi\sin\alpha\sin\delta+\cos\phi\cos\alpha\right)\\
s_z=\sin\delta\sin\zeta-\cos\zeta\sin\phi\cos\delta
\end{eqnarray}

The above set of equations toguether with $\vec{r}$ give us all the information
necessary to evaluate $\theta$. However, it is worth noting that, in the
determination of $\zeta$, the two valued arcsin provides us with 2 values for
the inclination angle: $\pm\zeta$ (i.e. we are not able to decide whether the
spin is pointing \textit{towards us} or \textit{away from us}).  To avoid this
problem we had to constrain our analysis to edge-on (or face-on) galaxies, i.e.
those with $\zeta\sim0^\circ$ (or $\zeta\sim90^\circ$). In practice, our sample
considered galaxies with b/a$\le$0.208  (or b/a$\ge$0.978), which implies that
$\zeta\le12^\circ$  (or $\zeta\ge78^\circ$).

 To achieve a sizeable sample of large voids we have used the Two-Degree Field
Galaxy Redshift Survey (2dFGRS; Colless et al. 2001) and the Sloan Digital Sky
Survey (SDSS; York et al. 2000). To characterize the uncertainty of the angle
$\theta$ we have used 2dFGRS and SDSS galaxy redshift survey 'mocks' constructed
from a set of large, high-resolution cosmological N-body simulations (Cole et
al. 1998). These simulations do not resolve detail on the scale of a galaxy so
we have assumed for this test that all the galaxies were edge-on with the
position angle of the rotation axis uniformly distributed in the sky plane.  We 
measured the difference between the angles $\theta$ estimated in real space and in
redshift space. We find that this difference is well described by a Gaussian
distribution with standard deviation of 10$^{\circ}$. In addition, these simulations
allow us to estimate the uncertainty at measuring the centre position of the
voids as $\pm$2.5 h$^{-1}$ Mpc. This uncertainty is smaller than the uncertainty in
estimating the position for individual galaxies: $\pm$4.2 h$^{-1}$ Mpc,  because the
central position of the void is determined using a large number of galaxies in
the shell. 

The 2dFGRS and SDSS (Data Release 3) catalogues give photometric and
spectroscopic information for hundreds of thousands of galaxies.  The 90\%
redshift completeness limit of the 2dFGRS is reached at b$_J$=19 mag and at
b$_J$=18.8 mag in SDSS (Norberg et al. 2002). If we wish to maximizing the
volume and the number of galaxies under exploration the above magnitude limits
imply that we must estimate the central position of the voids using all galaxies
brighter than M$_{bj}$=-19.32+5log(h) with z$<$0.14 in the 2dFGRS and z$<$0.13 in
the SDSS.  The number of voids that we detect with radius larger than 10
h$^{-1}$ Mpc is: 239 (2dFGRS) and 756 (SDSS).  The detection and size estimation
of the voids was performed with the HB void finder (Patiri et al. 2006).

 Once the largest voids are obtained we select all the edge-on and face-on
galaxies within the shells r$_{void}$$<$r$<$r$_{void}$+4 h$^{-1}$ Mpc
surrounding the voids. As mentioned before, we consider a spiral galaxy edge-on
if the ratio of semi-minor to semi-major axis is less than 0.208.  This
restriction removes from our sample spheroidal galaxies but leaves with a small
residue from close galaxy pairs that was eliminated by checking all the galaxies
individually by visual inspection.  Face-on galaxies are selected if the ratio
of semi-minor to semi-major axis is greater than 0.978. The face-on selected
sample is  contaminated by round spheroidal galaxies that are removed by visual
inspection. This last step can only be achieved in the SDSS since the 2dFGRS
survey lacks good quality image resolution allowing the visual classification of
the galaxies. Consequently, we only used edge-on galaxies in the 2dFGRS sample.

 To avoid edge biasing in our results we then removed all the above voids that
do not satisfy the condition that their radii plus the radius of the shells are
fully contained within the survey area. This leaves with 149 (2dFGRS) and 321
(SDSS) voids. The number of voids which contain at least one edge-on/face-on
spiral in their shells are 49 (2dFGRS) and 129 (SDSS).  The final number of
galaxies we analyse is 60 (edge-on; 2dFGRS), 118 (edge-on; SDSS) and 23
(face-on; SDSS). As expected by projection effects the number of face-on
galaxies is much smaller than the number of edge-on galaxies. The total number
objects in our sample is 201.  For each of these galaxies we derived the angle
$\theta$ between the rotation axis and the local void wall orientation as
explained above.

\section{Results and Discussion}

The observed probability density distribution of the  angle $\theta$ is shown in
Fig.  2.  Fig.  2a shows the result of a test placing the centres of the voids
randomly within the survey volume. The number of edge-on galaxies and their
distances to the centre of these voids are similar in this control experiment
than in the real case. As expected no signal is detected. When using the centres
of the real voids, both in the SDSS (Fig. 2b)  and 2dFGRS (Fig. 2c) surveys, the
distribution follows a similar trend, although the distribution for the 2dFGRS
is significantly noisier due to the smaller number of objects. The combined
sample is shown in Fig. 2d. The rotation axes of the galaxies tend to lie in the
plane parallel to the surface of the void (i.e. the  angles $\theta$ are
predominantly large).  We have used three different statistics to test the
rejection of the null hypothesis (i.e. that the orientation of the galaxies is
randomly distributed): the departure of the average (Avni \& Bahcall 1980) of
sin($\theta$) from 0.5 (i.e. the expected value in the null hypothesis case),
the Kolmogorov-Smirnov test and the $\chi^2$-test. The three statistics agree in
the rejection of the null hypothesis  at $\sim$99.7\% level ($\sin(\theta)$
test: 99.7\%; KS test: 99.6\% and $\chi^2$-test: 99.8\%) when the full galaxy
sample (Fig. 2d) is considered. For consistency with the 2dFGRS data we have
checked whether removing the face-on galaxies from the SDSS survey affects our
conclusion. As expected because of the small number of face-on galaxies the
signal only decreases very slightly ($\sim$0.1\%) when these objects are removed and
the main conclusion (i.e. the rejection of the null hypothesis) remains
unchanged. 

In Fig. 2d we show the comparison of our results with Lee's recent analytic
evaluation (Lee 2004) of the galaxy inclinations within the framework of the
tidal torque theory. The strength of the intrinsic galaxy alignment of the
galaxies with local shears at present epoch is expressed as the following
quadratic relation (Lee \& Pen 2002):
\begin{equation}
<L_iL_j>=\frac{1+c}{3}\delta_{ij}-c\widehat{T}_{ik}\widehat{T}_{kj},
\end{equation}
where $\bold{L}$ is the galaxy spin vector and $\bold{\widehat{T}}$ is the
rescaled traceless shear tensor $\bold{T}$  defined as
$\widehat{T}_{ij}$=$\widetilde{T}_{ij}/|\bold{\widetilde{T}}|$ with
$\widetilde{T}_{ij}$$\equiv$$T_{ij}-Tr(\bold{T})\delta_{ij}/3$, and c is a
correlation parameter introduced to quantify the strength of the intrinsic
shear-spin alignment in the range of [0,1].
The best agreement with the theory is obtained when Lee's
correlation parameter c is 0.7$^{+0.1}_{-0.2}$.  When c=1 the strength of the
galaxy alignment with the large-scale distribution is maximum, whereas c=0
implies galaxies are oriented randomly. It is worth noting that the value of c
measured in this work is a lower limit of the true value because it has been
evaluated without any attempt to correct for redshift distortion. Consequently,
the strength (and statistical significance) of the observed alignment in true
three dimensional space should be higher.  

Because of lack of resolution, using present-day N-body simulations what one can
predict is the intrinsic alignment of underlying dark halos rather than that of
the  observable luminous parts of the galaxies. Although the standard theory of
galaxy formation assumes that the rotation axes of the luminous parts align with
those of the dark halos, it was implied by recent hydrodynamical simulations
that the luminous components could conserve the initial memory of the intrinsic
shear-spin correlation better that the dark matter components (NAS04).
Interestingly,  our best-fit value of c=0.7 is $\ge$2 times higher than that
found in N-body simulations (Lee \& Pen 2000; Lee, Kang \& Jing 2005). This
reinforces the idea that the luminous parts of the galaxies tend to keep the
initial memory of the surrounding matter better than their dark matter
counterparts.

 Finally, the positive detection of alignments between galaxies and their
surrounding matter shown in this work would explain one of the  most intriguing
galaxy properties: the tendency of satellite galaxies to avoid orbits coplanar
with their host spiral galaxies (an effect known as the "Holmberg effect")
(Holmberg 1969; Zaritsky et al. 1997; Sales \& Lambas 2004). As we have shown,
the rotation axes of the galaxies preferentially lie on the sheets,
consequently,  the satellite galaxies that are accreted through the filaments
will populate mainly polar orbits.

\acknowledgments

We thank very useful discussions with Juan E. Betancort, Peter Coles,
Hans-Walter Rix, Robert Juncosa, Julio Navarro, Eric Bell, Jorge Pe$\tilde{n}$arrubia and
John Beckman.  Funding for the creation and distribution of the SDSS Archive has
been provided by the Alfred P. Sloan Foundation, the Participating Institutions,
the National Aeronautics and Space Administration, the National Science
Foundation, the U.S. Department of Energy, the Japanese Monbukagakusho, and the
Max Planck Society. The SDSS Web site is http://www.sdss.org/.  The SDSS is
managed by the Astrophysical Research Consortium (ARC) for the Participating
Institutions. The Participating Institutions are The University of Chicago,
Fermilab, the Institute for Advanced Study, the Japan Participation Group, The
Johns Hopkins University, the Korean Scientist Group, Los Alamos National
Laboratory, the Max-Planck-Institute for Astronomy (MPIA), the
Max-Planck-Institute for Astrophysics (MPA), New Mexico State University,
University of Pittsburgh, University of Portsmouth, Princeton University, the
United States Naval Observatory, and the University of Washington.


\clearpage

\begin{figure}
\epsscale{0.7}
\plotone{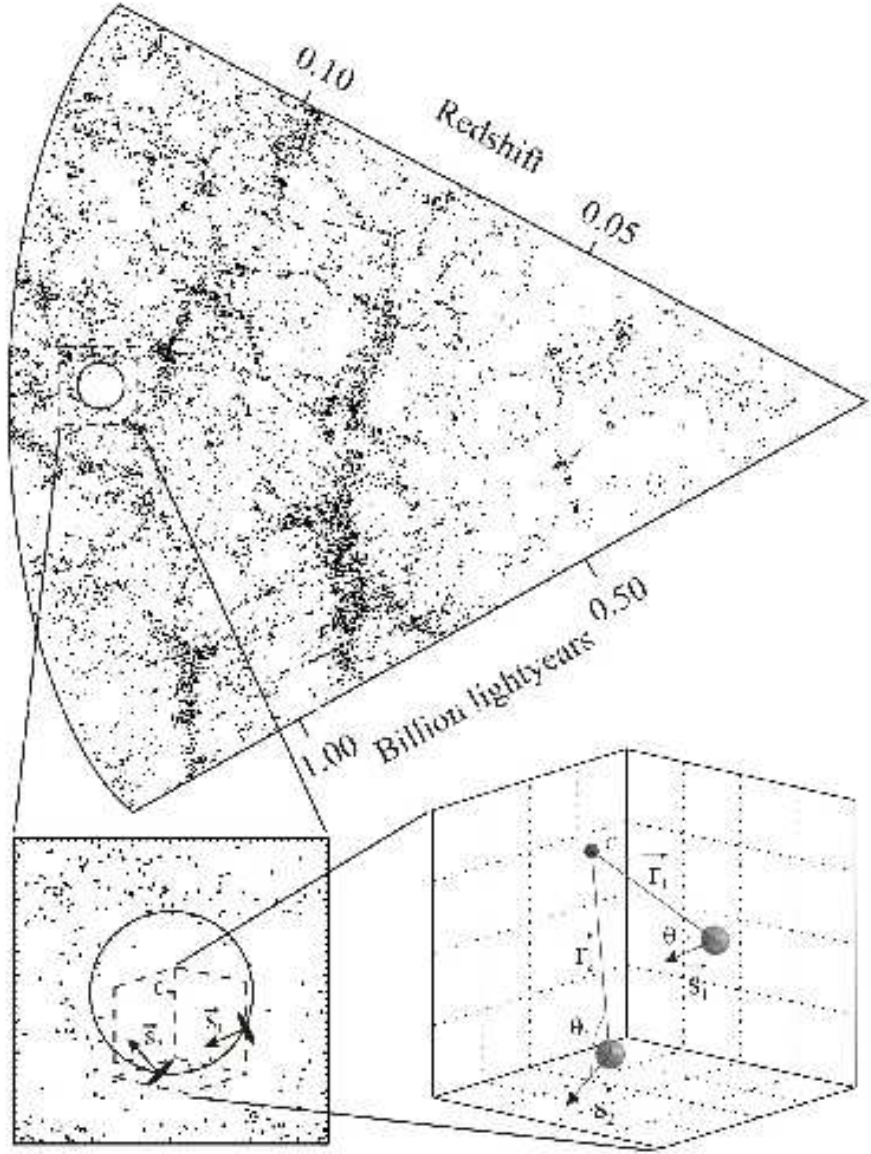}

\caption[]{Schematic representation of the technique used to
measure the angle $\theta$ between the spin axis of a galaxy $\vec{s}_i$ and the
orientation vector of the surrounding material. Large voids are detected in the
cosmic web.  We search for edge-on/face-on galaxies within the shells
surrounding the voids.  Because we select large voids,  $\vec{r}_i$ (the vector
joining the centre of the void to the centre of the galaxy) is a good
approximation to the vector normal to the shell of the void at the galaxy
position. Finally the angle $\theta$ between the above vectors is computed.}

\label{void} 
\end{figure}

\begin{figure}
\epsfig{file=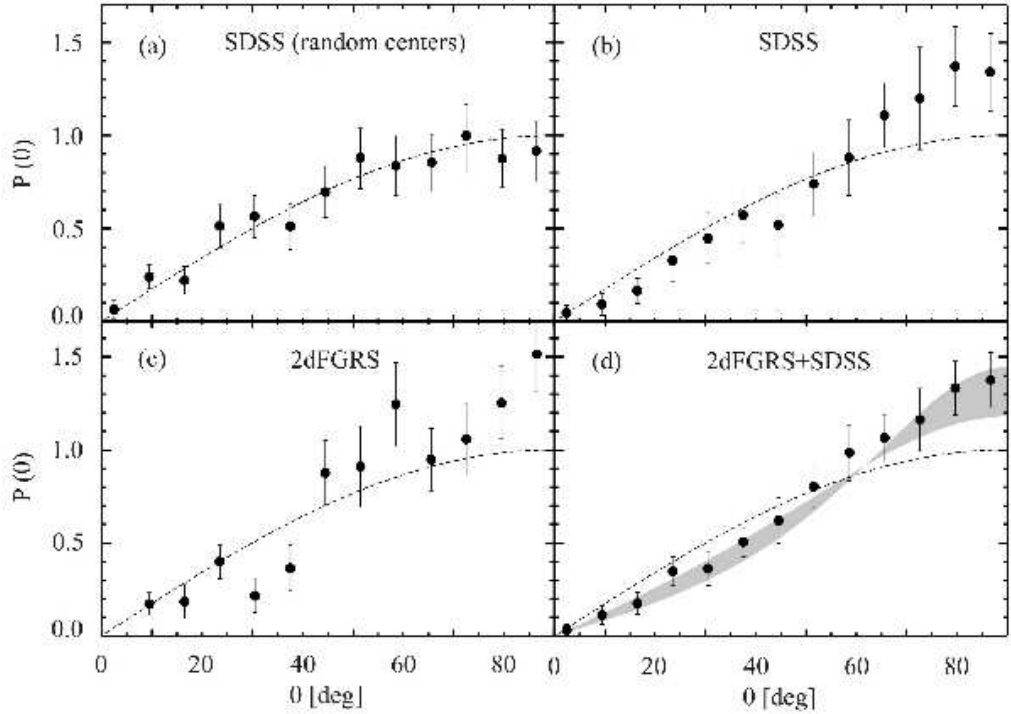,angle=0, width=\textwidth}
\vspace{-9cm}
\caption[]{Probability density distribution of the angles $\theta$ between the
rotation axes of the galaxies and the vectors normal to the planes defined by
their surrounding matter. The error bars on each bin represent the standard
deviation and are evaluated via 1000 Monte Carlo simulations where the
inclination of the galaxies was allowed to vary within a Gaussian distribution
with sigma equal to 6$^\circ$ (i.e. half the maximum angunlar inclination one of
our galaxies could have within our edge-on selection criteria, under the
assumption this galactic disks are infinitely thin). The error bars also include
the Poisson error.  The dashed line represents the null hypothesis (i.e. a sine
distribution). Panel (a) results from using random centres for the voids within
the survey volume. Panel (b) and (c) shows the output when the centres of the
large voids are used in SDSS and 2dFGRS. Panel (d) shows the angle distribution
when the full data set is used. The grey region in panel (d) corresponds to
Lee's analytic prediction (Lee 2004) for c=0.7$^{+0.1}_{-0.2}$. The c value is
estimated via a $\chi^2$ minimization.}

\label{statistics} 
\end{figure}

\end{document}